\documentstyle[aps,prl]{revtex}
\begin{document}
\draft
\title{Has the nonlinear Meissner effect been observed?}
\author{Klaus Halterman,$^{1}$ Oriol T. Valls,$^{1}$
 and Igor \v{Z}uti\'c$^{2}$ }
\address{$^1$Department of Physics and Minnesota Supercomputer
Institute, University of Minnesota, Minneapolis, Minnesota 55455}
\address{$^2$Department of Physics and Center for Superconductivity
Research,University of Maryland, College Park,
Maryland 20742}
\date{\today}
\maketitle
\begin{abstract}
We examine recent high-precision experimental
data on the magnetic field, ${\bf H}$, dependence of the penetration depth
$\lambda(H)$ in $\rm{YBa_2Cu_3O_{7-\delta}}$ (YBCO) for
several field directions in the $a-b$ plane.
In a new theoretical analysis that incorporates the effects of
orthorhombic symmetry, we show that 
the data at sufficiently high magnetic fields and
low temperatures are
in quantitative agreement with the theoretical predictions
of the nonlinear Meissner effect.
\end{abstract}
 
It is widely
accepted\cite{agl}  
that the symmetry of the order parameter
(OP) in high temperature superconductors (HTSC's) is at least predominantly
$d$-wave, vanishing at
nodal lines approximately ninety degrees apart in a quasi two-dimensional
Fermi surface (FS).
Many details of the OP in these materials remain quite unclear, however. Are
the nodes exactly at right angles? Are they true nodes or only very deep
minima, ``quasinodes''\cite{mix}?
Addressing these and similar questions is important for obtaining clues
about the nature of the superconductivity. They
are particularly
difficult to answer for the {\it bulk} OP, which may well differ \cite{bah}
from the more easily observed surface
state. To probe the bulk OP it is best
to use electromagnetic techniques, since electromagnetic fields
penetrate the sample over a depth of the order of the London                    
penetration length $\lambda$ which is, in these
materials, several orders of magnitude larger than the coherence length
$\xi$ which characterizes the range of typical surface probes. Indeed,
one of the early key results\cite{h1}
in support of bulk $d$-wave superconductivity
was the measurement of the linear
temperature dependence of the penetration depth
in a high-purity YBCO single crystal. However, such data indicated
only the existence of  nodal lines without the angular resolution needed
to identify their position. Consequently, intensive efforts to
precisely determine the structure of the bulk OP have continued.
 
It was first pointed out\cite{ys} years ago that
nodes in the OP yield distinctive and measurable
nonlinear effects in the
field and angular dependence of the
penetration depth when the superconductor is in the Meissner state.
In subsequent theoretical
work\cite{sv,ys2,zv1,zv2} more
emphasis was placed on the existence, due to this nonlinear Meissner
effect (NLME), of a  component of the diamagnetic moment
normal to the applied field, 
and on the torque associated with this transverse
component. These phenomena were deemed to be easier to
measure than the changes in $\lambda$ itself.
It was shown\cite{zv2} in this context that the NLME can be used to
perform {\it node spectroscopy}, that is, not just to infer the existence
of nodes, but to {\it locate their positions} on the FS, and to determine
whether they are true nodes or not.
Thus, the NLME is potentially a very important tool for the study of
the pairing state in HTSC 's, as well as other materials
in the ever increasing list of those for which the proposed OP leads
to an energy gap with nodes.
Yet, the experimental situation
is rather
confusing. The best experimental effort to measure the transverse
diamagnetic moment\cite{bhat}
in YBCO was inconclusive. Subsequently, results\cite{bid} for the magnetic
field
dependent penetration  depth to a precision\cite{bid2}
of $\sim 0.1$\AA \,  were reported. The NLME should be observable in such a high
precision experiment, more precise than 
existing transverse moment measurements. Unfortunately, no
theory of the NLME contribution to
the penetration depth for orthorhomobic structures such as YBCO
was available when Ref. \onlinecite{bid} was written. Only very
recently\cite{hvz} have the necessary calculations been performed.
This has resulted in
contradictory claims as to whether
observed results are in agreement or not
with  NLME theory.
Thus, a certain amount of
skepticism has developed as to the observability of the NLME.
 
In this paper we show that
measurements of the field
dependent penetration depth $\lambda(\psi,H)$ as
a function of the angle $\psi$ that an applied
field ${\bf H}$ in the $a-b$ plane
forms with the $a$-axis,
must be analyzed very carefully. The anisotropy of the linear
penetration depth tensor has a drastic effect\cite{hvz}
on the NLME  for $\lambda(\psi,H)$.
This may have been overlooked because the anisotropy effects
in the transverse moment are known \cite{zv2} to be relatively minor.
One must also  eliminate several other factors that may mask
the signal at low fields and which are very difficult to account for theoretically.
Thus we reanalyze here the best data available for the penetration depth
in YBCO. We find that, although some questions remain, the low
temperature data are
quantitatively in agreement with theoretical
expectations for the NLME in this material.

We  focus here on 
YBCO,
the most experimentally studied \cite{bhat,bid,ill,buan}
HTSC in this context. Hence, the relevant material
parameters are well known, thus reducing  the uncertainty in
the fitting procedures.
We perform our analysis primarily on the most complete available
high resolution data of Ref.~\onlinecite{bid} which includes results
for four different directions
of the applied field in the $a-b$ plane.
 
The angular and field dependent
increase in the penetration depth due to the NLME for materials
with orthorhombic anisotropy of the YBCO type was
first calculated  in Ref.~\onlinecite{hvz}.
The details will not be repeated here.
The sample is assumed to have its larger faces parallel to the $a-b$ plane
(this is the case for crystals grown by the usual methods)
and thickness large compared with the penetration depth. One  has
for the quantity
$\Delta \lambda(\psi,H)\equiv\lambda(\psi,H)-\lambda(\psi,0)$:
\begin{equation}
\label{calp}
\Delta\lambda(\psi,H) = \frac{1}{6} \frac{H}{H_0} \lambda {\cal Y}(\psi).
\end{equation}
Here $\lambda$ is the geometric mean of the 
two in-plane principal values,
$\lambda_a$ and $\lambda_b$, of the  zero field
penetration depth tensor,
$H_0$ is a characteristic field of order $\Phi_0/\pi^2\lambda\xi$     
($\Phi_0$ is the
flux quantum), and
${\cal Y}$ carries the angular dependence. The orthorhombicity, very
important in this case, is incorporated into ${\cal Y}$ through
two parameters\cite{hvz}: one is the ratio $\Lambda\equiv\lambda_a/\lambda_b$,
and the other is the angle $\alpha$ that the Fermi velocity at the node
located in the first quadrant
forms with the $a$ direction. Because of the orthorhombic distortion 
of the FS,
this angle does not have to  exactly equal $\pi/4$
even for a pure $d_{x^2-y^2}$ state, while
the quantity $\Lambda$, for YBCO, considerably exceeds unity.
Here we  take the zero field
quantities $\lambda_a$ and $\lambda_b$  fixed at their
experimental\cite{bas} values ($1050$\AA \, and $1575$\AA),
giving
$\Lambda=1.5$. In this
case, the full expressions\cite{hvz,cau}
for ${\cal Y}(\psi)$
simplify somewhat and can be written as:
\begin{mathletters}
\label{ybcowhole}
\begin{eqnarray}
\nonumber
{\cal{Y}}(\psi) &=& \frac{18 \Lambda}
{2+\Lambda}\cos^2\alpha\sin\alpha\cos\psi\sin^2\psi
+\frac{2}{\Lambda^2(1+2\Lambda)} \\ \nonumber
&\times& \sin^3\alpha\cos^3\psi
\Bigl[1+2\Lambda+
(4\Lambda-1)
\left(\frac{\tan\psi}{\tan\psi_1}\right)^{\frac{3\Lambda}
{\Lambda-1}}\,\Bigr] \\ \nonumber
&+& \frac{2\Lambda^2(2\Lambda^2-10\Lambda-1)}{(2+\Lambda)(1+2\Lambda)
}\cos^3\alpha\sin^3\psi
\left(\frac{\tan\psi}{\tan\psi_1}\right)^{\frac{3}{\Lambda -1}},\\ 
\psi &\in& [0,\psi_1],\\ \nonumber
{\cal{Y}}(\psi) &=& \frac{18}{1+2\Lambda}\sin^2\alpha\cos\alpha\cos^2\psi
\sin\psi +2\Lambda^2 \cos^3\alpha\sin^3\psi, \\
\psi&\in& [\psi_1,\frac{\pi}{2}].
\label{ybcowhole3}
\end{eqnarray}
\end{mathletters}
where the angle $\psi_1$ is given by 
$\psi_1 \equiv {\rm{arctan}}(\tan\alpha/\Lambda)$.
 
Because of the orthorhombicity, the angular dependence of
$\Delta\lambda(\psi,H)$ is quite different\cite{hvz}
for $\Lambda=1.5$ than that
found for the tetragonal case ($\Lambda=1$, $\alpha=\pi/4$). This is
unlike the situation for the transverse magnetic moment\cite{zv2},
which from symmetry considerations  vanishes when ${\bf H}$ is along
the $a$ or the $b$ axis. Thus,  moderate
orthorhombic anisotropy induces only  a relatively minor distortion
in the curve between $\psi=0$ and $\psi=\pi/2$ since these points are
so to speak, anchored. This is not the case for
$\Delta\lambda(\psi,H)$: when the field is applied along $\psi=0$
the currents flow over a region of thickness determined by $\lambda_b$
while if $\psi=\pi/2$ the relevant skin depth is $\lambda_a$. The effect
is nonzero, and different, in either case. This difference is
compounded by the nonlinearity and  the  apparent overall
symmetry of
${\cal Y}$  is $\pi$ rather than $\pi/2$ even for moderate
orthorhombicity. Failure to take this into account  leads
to erroneous conclusions concerning the angular dependence                      
of $\Delta\lambda(\psi,H)$.

To analyze data in terms of Eqs. (\ref{calp}) and (\ref{ybcowhole}),
additional
considerations are needed.
These expressions, indicating that 
$\Delta\lambda$  is proportional
to $H$, are valid at low temperature.  Here, ``low''
temperature is a {\it field dependent} concept. The characteristic
temperature separating the high and low $T$ regimes is
\cite{sv,ys2} $T^*(H)\approx
\Delta_0(H/H_0)$, where $\Delta_0$ is the  gap
amplitude.  At any finite $T$, the validity of the above equations
will break down
at sufficiently small $H$. Further, the effect of impurities is
not included. For the clean samples used in experiments\cite{bhat,bid,ill}
this should affect\cite{sv,ys2} only the small field results. The same
is true of possible nonlocal effects\cite{hirshy}.
If they are
present at all in this geometry\cite{kl},  they would affect
results at fields below\cite{com} 20 gauss. Ideally, one would like to take
into account all of these effects by modifying the above formulas. However,
it is not feasible at present to include 
all of these factors
{\it simultaneously} in a reliable manner. It is therefore best to perform
the analysis in a consistent manner in terms of data in the higher
range of fields available, where these additional effects are all weak.
 
In Fig.~\ref{fig1} we show best straight line fits to the 1.2 K data
of Ref.~\onlinecite{bid} for ${\bf H}$ along
the $a$ and $b$ directions. All data in
the range $H>60$ gauss are
included in the fit. The cutoff of 60 gauss was
chosen as the point below which deviations from a straight line begin and it
will be shown below to lead to a consistent interpretation. The straight
line does not intercept the origin of the original plot, which has to
be shifted downward.
This is as expected, since the
experimental $\Delta \lambda$ includes the previously mentioned
temperature\cite{shift},
impurity, and possible nonlocal effects which
increase this quantity with respect to the theoretical, clean, zero
temperature, local value. The shift is small, of order 1 {\AA}, indicating
that the sample is clean and any such spurious effects are small. The two
slopes of the lines
obtained from these fits are the quantities
${\cal Y}(0)\lambda/H_0$ and ${\cal Y}(\pi/2)\lambda/H_0$ respectively,
where from Eq.(\ref{ybcowhole}), 
${\cal{Y}}(0) = (2/\Lambda^2)\sin^3\alpha$, and 
${\cal{Y}}(\pi/2) = 2\Lambda^2\cos^3\alpha$.
Thus, the ratio
of the two slopes depends only on $\alpha$ since we have used the
independent experimental value
$\Lambda=1.5$.
We then determine the $\alpha$ that fits this  ratio and
subsequently
find the characteristic field $H_0$ from either one of the slopes.
The results are very sensible: we obtain $\alpha=\pi/4 + \pi/17$ and
$H_0=5660$ gauss. The value for the angle between the Fermi velocity
at the node and the $a$-axis
exceeds $\pi/4$ by a small amount, as one would expect
for a pure $d_{x^2-y^2}$ pairing state and a 
tight binding FS with a slight orthorhombic
distortion. The value of the characteristic field is
consistent with expectations\cite{zv1,zv2} and also
with our cutoff choice for the field: in the range of fitting
we have $H/H_0>0.01$. This means that in this field range, the
characteristic temperature $T^*(H)$ introduced above is of order of 4 to
12 K. Hence the 1.2 K data included in the fitting are 
in the low temperature regime and the procedure is consistent.
 
Up to now, we have, however, fit two quantities with two parameters,
although the reasonable values obtained for these parameters are
encouraging. To go beyond, we now use the obtained values of $\alpha$
and $H_0$ to plot the predicted slope of the high field data at
$|\psi|=\pi/4$
without any additional parameters. This is done in
Fig.~\ref{fig2}.
Experimental results for fields applied in the $\psi=\pm \pi/4$
directions are included. These results ought to be identical (even
with the orthorhombic distortion) and their small discrepancy reflects
systematic errors in the experiment. Nevertheless, the fit is excellent
in the high field range.
We also plot, (inset) with these parameter values, the predicted
angular dependence of $\Delta \lambda$ for YBCO. One can  see
that, with the orthorhombicity, this angular dependence
differs  considerably from that obtained for a tetragonal system, 
also  plotted for comparison. The actual curve is not symmetric
about $\pi/4$ and its maximum is much less pronounced than that for
the tetragonal case, which is characterized\cite{ys}
by a factor of $\sqrt 2$ between
maxima and minima. Because of these differences,
the
attempt made by the authors of Ref.~\onlinecite{bid} to reconcile the
angular dependence of their data with the theory of
the NLME in a tetragonal system, had to fail.
 
In Fig.~\ref{fig3} we compare the theoretical results with other
more recent data\cite{ill} on YBCO at $T$=1.4 K. The parameters used 
are exactly the
same as previously obtained. No new fits were performed. Results for the
two directions available (field applied along the two principal axes) are 
shown. This data is in a more restricted, lower field range, and it has
considerably more scatter than that of 
Bidinosti {\it et al}\cite{bid}. All that can be said with certainty
is that
it is  
consistent with the NLME theory with the same parameter
values.
 
In summary, the main result of the analysis presented here is that the
best low temperature, high field,
data\cite{bid} on the nonlinear penetration depth
in YBCO is in quantitative agreement, in its magnitude and angular and
and field dependence, with the NLME theoretical expectations.
Other data\cite{ill} are also consistent
with theory. Failure to observe the NLME in the transverse moment\cite{bhat}
seems to be attributable to the actual precision in that experiment
being just  slightly less than what was in fact required.
 
Two remarks
must be added: first, the crucial 
influence of the orthorhombic anisotropy
in the angular dependence of $\Delta\lambda$, which becomes very different
(see Fig.~\ref{fig2})
from that found for tetragonal symmetry,
must be emphasized. Second, one sees
the need to finesse the temperature, impurity and possibly other
problems associated with smaller fields by obtaining and
analyzing data at the highest possible fields below that of first
flux penetration. Fortunately, this field is in the range 200-400
gauss\cite{bhat,bid,buan}
for typical YBCO crystals.
 
The question of the temperature dependence of the results\cite{bid,ill}
is less clear and needs further discussion: results obtained at 7 K
for the same sample mainly discussed here are\cite{bid} not substantially
different from those at 1.7 K. With the characteristic temperature
$T^*(H)$ in the range estimated above, it can be that the high
field results are not yet affected by the
temperature at 7 K while those at low fields
are dominated by largely temperature-independent impurity effects. Indeed,
it appears that a straight line fit to the 7K data at the highest
fields (see Fig. 4 of Ref.~\onlinecite{bid})
has a larger (in absolute value) vertical axis intercept than that
for the 1.7 K data, which would be consistent with this scenario. Nevertheless,
the weak temperature dependence of the data will remain a puzzle so long
as a rigorous calculation including impurities, temperature, and possibly,
nonlocal and other effects is not feasible.
It is possible that these effects combine to yield a
temperature dependence weaker than what the naive theory would
predict.
 
Finally, our analysis indicates that there is no significant $is$
admixture to the $d_{x^2-y^2}$ gap, since such an admixture
would lead to quasinodes and to\cite{hvz}  a considerable reduction
in $\Delta\lambda$.  Furthermore,
the nearness of $\alpha$ to $\pi/4$ is consistent with the absence of
a real $s$ component as well.
 
It would be desirable to perform measurements of $\Delta \lambda$
in YBCO at additional values of the angle $\psi$ to verify in more
detail if the curve
in the inset of Fig.~\ref{fig2}
is indeed  closely followed.
The comments made here on the proper way to analyze experimental data
should also
be taken into account in any attempts to use the NLME to elucidate
the pairing states of other
suspected unconventional superconducting materials for which it is
estimated\cite{hvz} that 
the sensitivity of present  penetration depth measurements is sufficient
to probe the NLME.

We thank C.P. Bidinosti and A. Bhattacharya for very useful information about
their respective measurements. This work was
supported in part by Petroleum Research Fund, administered by the ACS (at
Minnesota)
and by DARPA and ONR (N00140010028) at Maryland.

\begin{figure}[t]
\caption{Magnetic field $(H)$ dependence of
$\Delta\lambda$ (see text). The straight lines are fits to
the 1.2 K data (circles and squares) of Fig. 3, Ref.~\protect\onlinecite{bid},
for $H>60$ gauss.
 Top:
$H$ applied along the $b-$axis.
 Bottom:
$H$ along the $a-$axis.}
\label{fig1}
\end{figure}

\begin{figure}[t]
\caption{$\Delta\lambda(H)$ for $H$ along $\psi = |\pi/4|$. 
The straight line is
the theoretical result at higher fields with the parameters
extracted from Fig. \ref{fig1}. Diamonds and triangles are the
experimental data with ${\bf H}$ applied
at $\psi = \pm \pi/4$. Inset:  predicted
angular dependence (thin curve) of $\Delta \lambda(\psi)$
including anisotropy.    
Bold curve:
result for a tetragonal system. The amplitudes of both curves
correspond to $H=180$ gauss.}
\label{fig2}
\end{figure}

\begin{figure}[t]
\caption{$\Delta\lambda$ as a function of H. The straight
lines are the theoretical results
with the same parameters
found in
Fig.~\ref{fig1}.  The symbols are experimental data of
Ref.~\protect\onlinecite{ill}.
Top panel:
${\bf H}$
at $\psi=\pm\pi/2$. Bottom panel: ${\bf H}$  along $\psi =0, \pi$. }
\label{fig3}
\end{figure}


\begin{references}
\bibitem{mix} These questions are sometimes phrased in terms
of the possible existence of an $s$ (for the first question) or $is$
(for the second) admixture to the predominant $d$-wave OP.
\bibitem{cau} The limit $\Lambda=1$ cannot be taken directly in Eq.~
\protect\ref{ybcowhole}. Recourse to the fuller expressions of
Ref.~\protect\onlinecite{hvz} is required.
\bibitem{shift} The temperature contribution to this
shift is seen in Fig.~14 of Ref.~\protect\onlinecite{ys2}.
\end{references}
\end{document}